\documentclass[preprint,pre,aps,showpacs]{revtex4}
\usepackage[latin1]{inputenc}
\usepackage{eufrak}
\usepackage{amsmath}
\usepackage{color}
\usepackage[dvips]{graphicx}
\usepackage{subfigure}

\newcommand{\mb}[1]{\mathbf{#1}}
\newcommand{\di}{\textrm{d}}
\newcommand{\eh}{\mathrm{e}}

\begin{document}

\title{Investigating quantum transport with an initial value representation of the semiclassical propagator}
\author{Christoph-Marian Goletz, Frank Grossmann}
\affiliation{Institut f\"ur Theoretische Physik Technische Universit\"at\\ 01062 Dresden, Germany}
\author{Steven Tomsovic}
\affiliation{Department of Physics and Astronomy\\ Washington State University\\ Pullman, Washington 99164-2814, USA}

\begin{abstract}
Quantized systems whose underlying classical dynamics possess an elaborate mixture of regular and chaotic motion can exhibit rather subtle long-time quantum transport phenomena.  In a short wavelength regime where semiclassical theories are most relevant, such transport phenomena, being quintessentially interference based, are difficult to understand with the system's specific long-time classical dynamics.  Fortunately, semiclassical methods applied to wave packet propagation can provide a natural approach to understanding the connections, even though they are known to break down progressively as time increases.  This is due to the fact that some long-time transport properties can be deduced from intermediate-time behavior.  Thus, these methods need only retain validity and be carried out on much shorter time scales than the transport phenomena themselves in order to be valuable.  The initial value representation of the semiclassical propagator of Herman and Kluk [M.~F.~Herman and E.~Kluk, Chem. Phys. \textbf{91}, 27 (1984)] is heavily used in a number of molecular and atomic physics contexts, and is of interest here.  It is known to be increasingly challenging to implement as the underlying classical chaos strengthens, and we ask whether it is possible to implement it well enough to extract the kind of intermediate time information that reflects wave packet localization at long times.  Using a system of two coupled quartic oscillators, we focus on the localizing effects of transport barriers formed by stable and unstable manifolds in the chaotic sea and show that these effects can be captured with the Herman-Kluk propagator.
\end{abstract}


\maketitle

\section{Introduction}

Model Hamiltonians for many molecular and atomic systems can be represented in the form of coupled anharmonic oscillators~\cite{Baer82, Child81, Frederick88}.  The dynamics of such systems, generally speaking, are not fully integrable, and may be said to be in the near-integrable regime if, for the most part, the resonances introduced into the dynamics by anharmonicities are non-overlapping~\cite{Chirikov59}.  On the other hand, if a great enough number of the resonances overlap strongly, a well developed domain of chaotic dynamics is generated, which coexists with other domains of near-integrable behavior.  In this case, the dynamics is in the mixed regime and the focus of the investigation presented here.  Due to the underlying dynamical diversity, quantized versions of such systems may display many remarkable properties and effects: for example, dynamical tunneling~\cite{Davis81, Child82}, chaos-assisted tunneling~\cite{Tomsovic94, Tomsovic98, Dembowski00} and ionization~\cite{Zakrzewski98}, and various forms of transport suppression and eigenstate localization~\cite{Vanleeuwen85, Casati86, Brown86, Geisel86, Jensen89, Galvez88, Mackay88, Bohigas93, Madronero06}.

Semiclassical theories, which date back to the early days of quantum mechanics~\cite{Jeffreys24, Wentzel26, Brillouin26, Kramers26, Vanvleck28}, provide a powerful approach to understanding the quantum dynamics of mixed systems, yet are incompletely developed.  Although trace formulae, which connect quantum densities of states to sums over periodic orbits, exist for integrable~\cite{Berry76,Berry77a}, near-integrable~\cite{Tomsovic95, Ullmo96}, and chaotic systems~\cite{Gutzwiller90}, analogous formulae are not known for mixed systems.  The study of dynamical tunneling in the presence of chaos still remains to be fully understood; see however~\cite{Shudo95, Creagh96, Shudo98,Tomsovic98b, Creagh99, Creagh99b, Creagh02, Brodier02, Eltschka05, Baecker08} for progress.  Similarly, the time-dependent or dynamical basis of various mechanisms of eigenstate localization are poorly understood, and there has not been a consensus that such localizing effects are captured in a full semiclassical approach~\cite{Yoshida04,Kaplan98,Maitra00}.  The discovery of a sum rule for subtle classical orbit action correlations seems to support the studies which find the semiclassical theories capable of describing aspects of localization~\cite{Sieber01,Richter02}.

However, it is not necessary to resolve this question to deduce useful information.  For example, Bohigas et al.~\cite{Bohigas93} showed that in the presence of classical transport barriers (for example, cantori) intermediate time scale dynamics are sufficient to deduce aspects of long-time dynamical properties such as suppression of transport.  It is known that even for strongly chaotic systems, non-uniformized implementations of semiclassical dynamics are accurate to time scales that lengthen much more in the short wavelength limit than the time for chaos to develop on an ever finer scale~\cite{Oconnor92, Tomsovic93}.  This is usually expressed in terms of a scaling with Planck's constant $h$.  For smooth Hamiltonians, such as strongly coupled-anharmonic oscillators, the time scale of accurate semiclassical propagation would lengthen algebraically as $h^{-1/3}$~\cite{Sepulveda92} or better.  For integrable systems, the expectation is at least as long as $h^{-1}$.  Mixed systems can be made to work at least as well as chaotic systems.  The time scale for chaos to develop to the course-grained volume represented by a quantum state, roughly $h^d$ (where $d$ is the number of degrees of freedom), only grows logarithmically as $\tau\propto \mu^{-1} \ln h^{-1}$, where $\mu$ is the Kolmogorov-Sinai entropy (essentially related to positive Lyapunov exponents).  So although transport suppression (or localization) time scales would roughly be much, much longer, $\propto h^{-d}$, the dynamics between the logarithmic and relevant algebraic time scales does provide a window wherein interference is dominating the dynamics and with which there is at least an opportunity to deduce something about long-time properties such as quantum transport suppression or eigenstate localization.

In the phase space region dominated by chaotic dynamics, it is quite common to find partial barriers that limit transport connections between subregions of the chaotic domain.  The barriers can be constructed as surfaces having the property that locally a minimum flow of phase points cross from one side to the other per unit time; i.e. a minimized flux.  These bounding surfaces can take the form of so-called cantori~\cite{Mackay84a, Mackay84b} or intersections of stable and unstable manifolds of short periodic orbits~\cite{Channon80, Bensimon84, Mackay87}, both of which in two-degrees-of-freedom are associated with turnstiles that determine quantitatively how impeded the flow is.  They introduce additional time scales into the dynamics.  The semiclassical theory must be developed and valid long enough to be able to deduce these time scales.

Semiclassical theories can be based on stationary phase approximations applied to various representations of the Feynman path integral.  In atomic, chemical, and molecular physics applications, the so-called initial value methods~\cite{Heller81, Herman84, Kay94} are heavily relied upon, more so than the boundary value methods, such as in the work of van Vleck~\cite{Vanvleck28}.  It is of interest to understand better the limits and strengths of these initial value methods.  Here, we focus on the implementation derived by Herman and Kluk (HK)~\cite{Herman84} which was shown to be very efficient.  The dynamics of molecules and atoms can be treated successfully with this approach \cite{Kay94b,Grossmann99,Thoss04,Kay05}, although the more unstable the dynamics in any chaotic domain, the more challenging the work.  Here, we apply it to the semiclassical propagation of wave packets in chaotic regions of a two-dimensional mixed phase space quantum system.  It turns out that the HK method is suitable for propagation times where localization due to transport barriers becomes manifest.  In contrast to previous works of Yoshida et al.\ \cite{Yoshida04} and Maitra \cite{Maitra00}, we focus on obtaining the critical intermediate-time dynamical information of a mixed system in a parameter regime where the chaotic domain is strongly unstable, yet nevertheless retains significant partial transport barriers.

This paper is structured as follows: in Sect.~\ref{sec-2} we review using the propagation of localized wave packets as a means of investigating the suppression of quantum transport due to partial classical transport barriers in phase space.  Intermediate propagation time dynamics is connected to suppression of quantum transport seen at long times with a combination of semiclassical, random matrix ensemble, and master equation approaches.  This is followed by a description of the semiclassical initial value representation and purely classical methods in Sect.~\ref{sec-3}.  The coupled quartic oscillator model system is introduced in Sect.~\ref{sec-4}. In Sect.~\ref{sec-5} our numerical results are presented and followed with a brief discussion and conclusion.

\section{Inhibiting quantum transport}
\label{sec-2}

Because it has not received a great deal of attention and it is central to our work, we review in this section previous work demonstrating the connection between short- and intermediate-time propagating wave packet behaviors and long-time, inhibited quantum transport due to partial classical transport barriers~\cite{Bohigas93,Smilansky92}.  Of course, the long-time suppression of the spreading of a wave packet implies a localization of the eigenfunctions of the system and vice versa so that these two topics are manifestations of the same physics and discussion of one has its translation for the other.  The weakest forms of localization related to the chaotic zones in phase space, i.e.~coherent backscattering/weak localization~\cite{Baranger93} and eigenstate scarring by short, unstable periodic orbits~\cite{Heller84}, are specifically excluded from the approach presented in this section through the assumptions regarding local mixing and the absence of certain correlations.  However, they have been discussed at length elsewhere, would increase the localization discussed below, but are not of primary importance here. 

\subsection {Transport barriers' connection to quantum states}

Consider a system with a mixed classical dynamics such that on any energy surface of interest, there exist transport barriers which can be used to separate the full chaotic phase space part into several subregions.  Following Bohigas et al.~\cite{Bohigas93}, a first critical concept in relating these classical transport barriers to quantum behaviors is defining state vector subspaces associated with each separate region of the available chaotic phase space in a narrow energy range of interest.  It is with respect to any basis, which respects these subregions, that gives one a foundation for discussing eigenfunction localization in the dynamical context.  There are a few parameters containing the most critical information, i.e.~the $j^{th}$ subregion's phase space volume ${\cal V}_j$ in a narrow (Thouless) energy window $\Delta E$~\cite{Ullmo08}, vector subspace dimensionality $N_j\approx {\cal V}_j/h^d$, relative fraction of phase space volume $f_j= {\cal V}_j/{\cal V}$, and a shortest mixing time scale $\tau_m \propto \mu^{-1} \ln \left({\cal V}_j/ h^d\right)$.  Assuming all the subregions' volumes are the same order of magnitude, $\tau_m$ is the time for the chaotic mixing in the dynamics to take a cell of volume $h^d$ to mix into all the cells within one of these chaotic subregions.  A transport barrier is relevant if its flux (defined on a fixed energy surface) crossing from the $j^{th}$ region into the $k^{th}$ region, $\phi_{jk}$, is small enough that ${\cal V}_j/\phi_{jk}>> \tau_m \Delta E$; generally $\phi_{jk}$ is energy dependent, i.e.~depends on the local energy surface in question.  From this, one sees already that in the short wavelength limit (summarized as $h\rightarrow 0$), any single barrier loses its ability to be responsible for localizing effects with respect to the defined subregions because in this limit eventually $\tau_m \Delta E >> {\cal V}_j/\phi_{jk}$.  However, this limit is approached extremely slowly due to the logarithmic dependence.  In a broad range of applications, transport barrier time scales are found to be quite relevant.  In the case of diffusive dynamics, an infinity of barriers may be present, each of which is nearly negligible.  This is intuitively a dynamical basis for Anderson localization~\cite{Anderson58,Fishman82}.

It is the mixing time scale that underlies the possibility of a combined semiclassical and statistical approach.  In the vector subspace associated to a single chaotic subregion, quantum states that are approximately stationary (up to times of ${\cal V}_j/[\phi_{jk}\Delta E]$) have the statistical properties implied by random matrix theory according to the Bohigas-Giannoni-Schmit conjecture~\cite{Bohigas84} (random matrix theory provides a proper way to define ensembles with which to define statistical measures).  Crudely, these states can be thought of as ``unperturbed'' eigenstates of a system in which the transport fluxes crossing from one region to another are vanishing.  Thus, the quantum evolution operator is described by a matrix that divides into $n \times n$ blocks, where $n$ is the number of subregions.  Each diagonal block would have the properties of the unitary random matrix ensemble if all the fluxes $\phi_{jk}$ vanished.  In that case, except for tunneling tails, any initial wave packet essentially within one subregion would remain within its respective subspace for all propagation times, and the system's Hamiltonian expressed in this representation would also have diagonal blocks; its spectral statistical properties would mimic a weighted random superposition of random matrix spectra.  For non-vanishing $\phi_{jk}$, the evolution operator and the Hamiltonian would have matrix elements connecting the diagonal blocks, and thus a wave packet would leak out at least partially from the subspace to which it initially belonged.  In French et al.~\cite{French88a}, it was stressed that there existed a universal, dimensionless transition parameter $\Lambda$ for symmetry breaking problems (here a dynamical symmetry is being broken) defined as the mean square admixing matrix element of the Hamiltonian $v_{jk}^2$ divided by the mean level spacing squared $D^2$.  They emphasized that although it is straightforward to identify the transition parameter through perturbation theory, it had a far more general meaning and contained the necessary information well beyond the validity of perturbation theory.  For this problem, Bohigas et al.~\cite{Bohigas93} showed semiclassically that the quantum transition parameter could be expressed in terms of the classical fluxes and relative fractions of phase space volumes through the relation
\begin{equation}
\Lambda_{jk}\equiv \frac{v_{jk}^2}{D^2}=\frac{\phi_{jk}}{4\pi^2 h^{d-1}f_j f_k},
\label{lambda}
\end{equation}
and it determines the degree to which dynamical localization is present in the system.  For very small values of the transport flux relative to $h$, 
presumably this formula should be modified to incorporate a tunneling contribution. However, to our knowledge the tunneling correction is still an unsolved problem. 

It is worth noting here a relationship between a quantum transition probability and the transition parameter.  In a perturbative regime, for short times (but still longer than the logarithmic mixing time scales), and a single barrier, the relationship is given by~\cite{Bohigas93}
\begin{equation}
\langle \hat{U}^\dagger(t)\hat{P}_j \hat{U}(t)\hat{P}_k \rangle_E= \frac{2\pi}{\hbar \bar{\rho}(E)} f_j f_k \Lambda_{jk}t.
\label{ampli}
\end{equation}
where $\bar{\rho}(E)$ is the local mean level density, $\hat{U}(t)$ is the unitary evolution propagator, and the $\{\hat{P}_j,\hat{P}_k\}$ are projectors into the subspaces defined above;  as indicated, the expectation value is taken locally in energy.  The quantity on the l.h.s. above measures how much of the $k^{th}$ subspace leaks into the $j^{th}$ subspace under evolution.  This can be envisioned as a kind of Fermi-Golden-Rule relation for chaotic systems.

\subsection{A measure of quantum transport suppression}
\label{alm}

In order to investigate quantum transport using the dynamics of wavefunctions, it is convenient to consider wave packets $\Psi_j$ which reside within some particular vector subspace associated with a (say the j$^{th}$) phase space region.  The propagations of these initially ``localized'' wave packets $\Psi_j(t)$ are then studied by observing their projections onto the various other localized wave packets as a function of time.  A particularly convenient measure is given by

\begin{equation}
{\cal P}_{jk}=\lim\limits_{\Delta T \rightarrow\infty}\frac{1}{\Delta T}\int\limits_{T}
^{T+\Delta T}\left| \left\langle \Psi_k | \Psi_j (t) \right\rangle \right|^2 \di t,
\label{Delta_orig}
\end{equation}
which is the long time average of the absolute square of $\Psi_j(t)$ projected onto some $\Psi_k$.  The notation $\langle \cdot \rangle$ indicates the appropriate norm (overlap).  Assuming the partial barriers are effective, $\Psi_j(t)$ will spread with different mean intensities into each region, which will saturate at some point and not increase any further no matter how long a propagation time considered.  Statistically speaking, it is possible to deduce the extent of such a localization property from the ${\cal P}_{jk}$ with one extra ingredient, i.e. the smoothed local spectra of $\Psi_j$ (and $\Psi_k$),
\begin{equation}
\label{localspectra}
\bar{S}_j (E) = \sum_{l=1}^{N_j} \left| \langle l | \Psi_j \rangle \right|^2 \delta_\epsilon \left( E -E_n \right)
\end{equation}
where the $\{| l \rangle\}$ are the previously mentioned approximately stationary states of the $j^{th}$ region (unperturbed eigenstates) and $\epsilon$ indicates that the $\delta$-functions are smeared sufficiently to smooth out oscillations locally in energy.  For a Gaussian wave packet not-too-close to a turning point, there is an approximate analytic expression for the envelope of the local spectrum given by Fourier transforming the initial decay of the autocorrelation function~\cite{Heller81}, which gives, 
\begin{equation}
\bar{S}_l(E) \simeq \frac{1}{\sqrt{\pi\gamma}\hbar\|\mb{p}_0 \|} \exp \left(-\frac{(E-E_0)^2}{\gamma\hbar^2\mb{p}_0^2} \right).
\label{S}
\end{equation}
Here $\gamma$ is a width parameter, $\mb{p}_0$ the mean momentum, and $E_0$ the mean energy.   If the wave packet is close to a turning point, a more accurate expression should be used. 

It was shown~\cite{Bohigas93} that
\begin{eqnarray}
{\cal P}_{jk} & = & \frac{\Delta_{jk}({\bf \Lambda})}{ f_k\bar{\rho}_c} \int \di E\ \bar{S}_j (E) \bar{S}_k (E) 
\label{C_pre} \nonumber \\
& \simeq & \frac{\Delta_{jk}({\bf \Lambda})}{ f_k\bar{\rho}_{c}} \frac{1}{\sqrt{\pi}\hbar \sqrt{\mb{p}_{0,k}^2\gamma_k+\mb{p}_{0,j}^2\gamma_j}}
\end{eqnarray}
where $\bar{\rho}_c$ is the locally smoothed density of states for the full chaotic region.  All the information about the extent of localization is contained in the function $\Delta_{jk}({\bf \Lambda})$ whose meaning is given by the projection into the $k^{th}$ subspace of a propagated approximately stationary state from the $j^{th}$ region
\begin{equation}
\label{deltaprojection}
\Delta_{jk}({\bf \Lambda}) = \lim\limits_{\Delta T \rightarrow\infty}\frac{1}{\Delta T}\int\limits_{T}
^{T+\Delta T} \di t\ \left| \left|\hat P_k | l (t) \rangle \right| \right|^2 \; .
\end{equation}
In a first order perturbative regime (i.e.~small transport fluxes regime), it can be derived analytically with the result~\cite{Bohigas93}
\begin{eqnarray}
\Delta_{jk}({\bf \Lambda}) &=& f_k\sqrt{2\pi\Lambda_{jk}} \qquad \qquad\quad (j\neq k) \nonumber \\
\Delta_{jj}({\bf \Lambda}) &=& 1- \sum_k f_k\sqrt{2\pi\Lambda_{jk}}
\label{Delta}
\end{eqnarray}
and in the opposite extreme of very strong mixing, it's value saturates at 
\begin{equation}
\lim_{\bf \Lambda \rightarrow \infty} \Delta_{jk}({\bf \Lambda}) \Rightarrow f_k \; .
\end{equation}
The closer $\Delta_{jk}({\bf \Lambda})$ is to the value $f_k$, the less localization in the eigenstates or the less suppression of quantum transport.

The perturbation result for $\Delta_{jk}({\bf \Lambda})$ can be generalized using a Master equation approach developed in Smilansky et al~\cite{Smilansky92}.  It relies on the construction of a Markov matrix whose eigenvalues and eigenvectors contain the all the information about transport connections and volumes of the subregions.  They define a classical probability vector $\bf P$ with $n$ components that at any moment in time gives the probability of being found in one of the $n$ subregions.  Defining the transition probability matrix $\bf A$, whose elements are given by
\begin{eqnarray}
A_{jk} &=& \frac{-\phi_{jk}}{{\cal V}_j} \qquad \qquad j\ne k \nonumber \\
A_{jj} &=& \sum_{{k=1\atop (\ne j)}}^n \frac{\phi_{jk}}{{\cal V}_j}
\end{eqnarray}
gives the probability vector at any time
\begin{equation}
{\bf P}(t) = \exp(-{\bf A} t) {\bf P}(0)
\end{equation}
Since total probability is preserved, there is always one vanishing eigenvalue of $\bf A$ whose corresponding eigenvector represents the state of complete mixing or uniform probability of being found anywhere in the volume.  The rest of the eigenvalues and eigenvectors describe the decay of any initial probability towards the state of complete mixing.  The smallest eigenvalue bounds the time it takes for this to occur.  The quantum spectral fluctuations were related to the trace of the operator $\exp(-{\bf A}t)$ using a semiclassical theory.  A similar calculation could be done for the $\Delta_{jk}({\bf \Lambda})$.

\section{The Semiclassical Propagator in Initial Value Representation}
\label{sec-3}

For the semiclassical dynamic results presented ahead, an implementation of the HK propagator~\cite{Herman84} is used.  It is reviewed below as is the linearized semiclassical initial value representation (LSC-IVR)~\cite{Sun98,Wang98}, which served as our workhorse to produce purely classical results to compare with the semiclassical and exact quantum ones.

\subsection{The Herman-Kluk Propagator}

The HK propagator is frequently used in chemical and molecular physics when treating systems with a large number of degrees of freedom numerically. Its origins go back to the work of Heller on frozen Gaussian wave packet dynamics~\cite{Heller81} and as an initial value representation, no root searches for classical trajectories need to be performed, as is the case for the Van Vleck-Gutzwiller (VVG) propagator~\cite{Vanvleck28,Gutzwiller90}. This fact may be advantageous especially for systems with several degrees of freedom, where root searching may become cumbersome.  Formally, however, the HK is closely related to the VVG propagator via a stationary phase approximation~\cite{Grossmann99}.

For a $d$-degree-of-freedom system, the HK propagator reads
\begin{equation}
K^{\mathrm{HK}}(\mb{x},t;\mb{x'},0) =\iint \frac{\di^d q_0\, \di^d p_0}{(2\pi\hbar)^d}\; \langle \mb{x}|g_\gamma(\mb{q}_t,\mb{p}_t)\rangle \cdot R(\mb{q}_0, \mb{p}_0,t)\;\eh^{\imath S/\hbar }\; \langle g_\gamma(\mb{q}_0,\mb{p}_0)|\mb{x}'\rangle.
\label{hk}
\end{equation}
Its basic ingredients are coherent states given in the position representation
\begin{equation}
\langle \mb{x}|g_\gamma(\mb{q},\mb{p})\rangle=\left( \frac{\gamma}{\pi} \right)^{d/4} \eh^{-\gamma(\mb{x}-\mb{q})^2/2 + \imath \mb{p} \cdot(\mb{x}-\mb{q})/\hbar},
\end{equation}
with the $d\times d$ width parameter matrix $\gamma$, which we assume to be diagonal and equal for $\langle \mb{x}|g_\gamma(\mb{q}_t,\mb{p}_t) \rangle$ and $\langle g_\gamma(\mb{q}_0, \mb{p}_0) | \mb{x}' \rangle$.  Classical dynamics enters the expression through the propagation of trajectories from $(\mb{q}_0,\mb{p}_0)$ to $(\mb{q}_t,\mb{p}_t)$ using Hamilton's equations, and the calculation of the classical action $S=\int_0^t {\cal L}\, \di t'$, with $\cal L$ being the Lagrangian.  An important role is played by the prefactor
\begin{equation}
R(\mb{q}_0, \mb{p}_0,t)=\frac{1}{2}\sqrt{\det\left(\mb{\underline{m}}_{11}+\mb{\underline{m}}_{22}-\imath \frac{\gamma}{\hbar}\mb{\underline{m}}_{21}+\frac{\imath\hbar}{\gamma}\mb{\underline{m}}_{12}\right)},
\end{equation}
with the elements of the stability (or monodromy) matrix:
\begin{equation}
\mb{\underline{M}}=
\begin{pmatrix}
\mb{\underline{m}}_{11} & \mb{\underline{m}}_{12} \\
\mb{\underline{m}}_{21} & \mb{\underline{m}}_{22}
\end{pmatrix}
=
\begin{pmatrix}
\frac{\partial\mb{p}_t}{\partial \mb{p}'} & \frac{\partial\mb{p}_t}{\partial \mb{q}'} \\
\frac{\partial\mb{q}_t}{\partial \mb{p}'} & \frac{\partial\mb{q}_t}{\partial \mb{q}'}
\end{pmatrix}
\end{equation}
For Gaussian wave packets, correlation functions can be computed directly with the HK propagator, according to
\begin{eqnarray}
c_{jk}(t) & = & \int \di^d x\; \Psi_k^*(\mb{x})\Psi_j(\mb{x},t) \nonumber \\
& = & \iint \frac{\di^d q_0\, \di^d p_0}{(2\pi\hbar)^d}\;\langle \Psi_k(t_0)|g_\gamma(\mb{q}_t,\mb{p}_t) \rangle\; R(\mb{q}_0, \mb{p}_0,t) \eh^{\imath S/\hbar}\; \langle g_\gamma(\mb{q}_0,\mb{p}_0) | \Psi_j(t_0) \rangle,
\end{eqnarray}
thus avoiding explicit calculation of the wave function.  A $2d$-dimensional phase space integral remains which is usually calculated via a Monte-Carlo integration~\cite{Metropolis49}.

It is worthwhile to mention that the HK propagator is
the lowest order term in a series 
expression for the full quantum mechanical propagator~\cite{Zhang03,Kay06}.
Considering higher order terms in this series, the description of 
(deep) barrier tunneling is within the reach of semiclassical IVR 
methods~\cite{Hochman06,Hochman08}. The numerical effort to determine the
correction terms is quite formidable, however. 

\subsection{The Linearized Semiclassical Initial Value Representation}

The LSC-IVR serves to produce classical results and is summarized very briefly. It was developed by Miller and co-workers as an approximation to semiclassical IVRs, avoiding oscillating integrands~\cite{Sun98, Wang98}.  Starting with an expression for the transition probability, i.e.~the squared absolute value of the correlation function, utilizing a van-Vleck-type IVR~\cite{Miller01}, and linearizing the phase differences of the trajectories one finally ends up with
\begin{equation}
P_{jk}(t) \equiv |c_{jk}(t)|^2 = \frac{1}{(2\pi\hbar)^d}\iint \di^d q_0 \di^d p_0 \, A_{(w)}(\mb{q}_0,\mb{p}_0) B_{(w)}(\mb{q}_t,\mb{p}_t),
\end{equation}
where in this case $A_{(w)}$ is the Wigner-Weyl transform of the associated initial wave packet
\begin{equation}
A_{(w)}(\mb{q}_0,\mb{p}_0)=\int \di^d \tilde{q}\, \eh^{-(\imath \mb{p}_0\cdot\tilde{\mb{q}})/\hbar} \langle \mb{q}_0+\frac{\tilde{\mb{q}}}{2}|\Psi_j\rangle \langle \Psi_j|\mb{q}_0-\frac{\tilde{\mb{q}}}{2}\rangle.
\end{equation}
Similarly, $B_{(w)}(\mb{q}_t,\mb{p}_t)$ is defined using the final wave packet. Inserting explicit expressions for the wave packets $\Psi$ gives
\begin{multline}
P_{jk}(t)=\left(\frac{2}{\pi\hbar}\right)^d \iint \di^d q_0 \di^d p_0 \exp\bigg[-\frac{1}{\gamma\hbar^2}(\mb{p}_0-\mb{p}_j)^2 \\
-\gamma (\mb{q}_0-\mb{q}_j)^2 -\frac{1}{\gamma\hbar^2}(\mb{p}_t-\mb{p}_k)^2-\gamma (\mb{q}_t-\mb{q}_k)^2 \bigg],
\label{classical}
\end{multline}
which is much easier to apply than the HK method.  However, all phase information of the trajectories is lost and therefore Eq.~(\ref{classical}) represents a purely classical result.

\section{The model system}
\label{sec-4}

As a model system for calculations, consider the two-dimensional quartic oscillator whose Hamiltonian is given by~\cite{Bohigas93}
\begin{equation}
H(\mb{p},\mb{q})=\frac{p_1^2+p_2^2}{2}+ a\left(\frac{q_1^4}{b}+bq_2^4+2\lambda q_1^2 q_2^2\right) ,
\end{equation}
where $\lambda$ is the coupling coefficient and $a=0.2$ a scaling factor. The factor $b$ allows one to tune the symmetry of the system. We use $b=\pi/4$ to have the convenient rectangular symmetry and secondly to be sufficiently far away from a ``stripe'' symmetry, which would involve a quasi one-dimensional system.  This system is quite suitable because it reveals all kinds of dynamical behavior depending on the value of $\lambda$.  It can be tuned continuously from the fully integrable case via the mixed case towards the fully chaotic case in the sense, that nearly all trajectories explore the entire phase space for arbitrarily long times.
\begin{figure}
\includegraphics[width=0.5\textwidth]{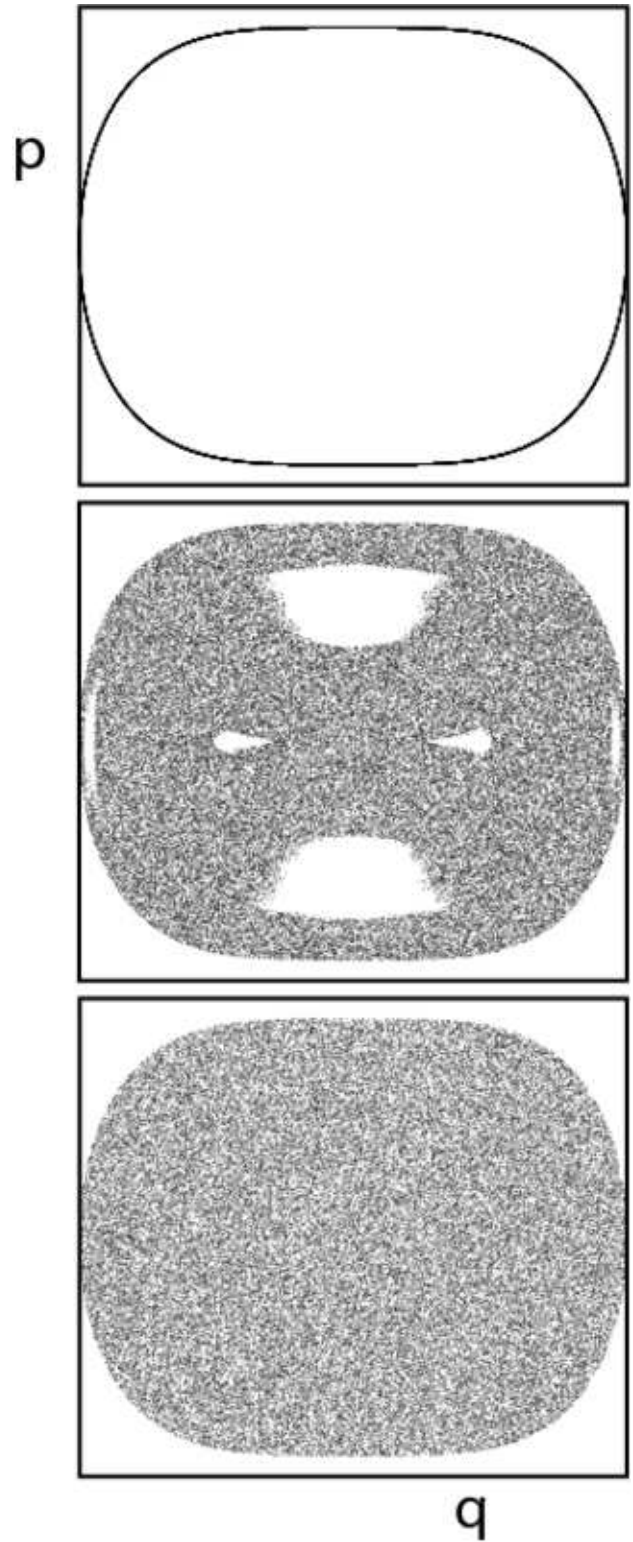}
\caption{$(q_1,p_1)$-Poincare sections illustrating the chaotic domain for the coupling coefficients $\lambda=0.0$, $\lambda=-0.35$ and $\lambda=-0.6$ from top to bottom respectively.  Reprinted from~\cite{Bohigas93} with permission.}
\label{poincare}
\end{figure}

\begin{figure}
\includegraphics[width=0.5\textwidth]{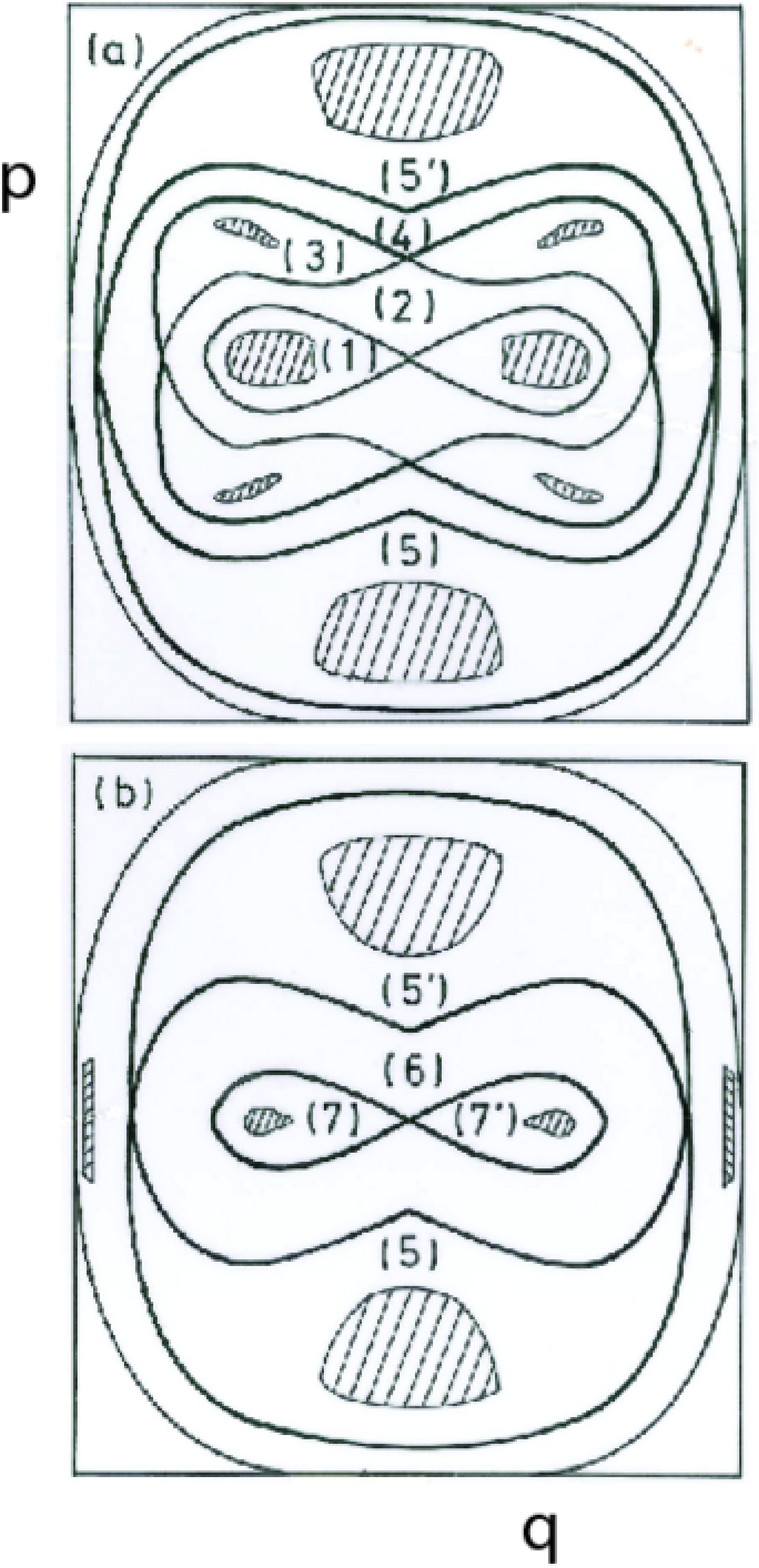}
\caption{$(q_2,p_2)$ and $(q_1,p_1)$-Poincare sections for $\lambda=-0.35$ showing the different subregions in the chaotic sea and the regular islands; the large regular island inside region $5^\prime$ is denoted as island $\mathcal{A}$; modified from~\cite{Bohigas93} and reprinted with permission.}
\label{regions}
\end{figure}

For the visualization of a 4-dimensional phase space, a Poincar\'e section (or surface of section) is a standard and convenient tool. For each of the three dynamical cases shown in Fig.~\ref{poincare}, the $(q_1,p_1)$-Poincare sections ($p_2>0$) are for a single trajectory. The first case shows regular dynamics and corresponds to the situation of two uncoupled quartic oscillators. The regular motion (for other individual invariant tori) would appear as concentric curves around the origin.  As $\lambda$ is decreased, the dynamics become mixed with near-integrable islands (whose dynamics are governed by the KAM-theorem) surrounded by a chaotic sea. Arriving at $\lambda=-0.6$, the whole phase space can be regarded as being fully chaotic ($99.5\%$ of the phase space is chaotic).  A further decrease to $\lambda=-1.0$ does not change the dynamics dramatically with the volume of the chaotic region being virtually identical with the phase space volume. This is also the borderline case which separates the bounded from the unbounded system and it is an interesting example of a system with a non-compact phase space, but one which nevertheless has only a discrete spectrum~\cite{Tomsovic91}.  

The case of greatest interest in this work is the mixed case with $\lambda=-0.35$.  For this system, the various chaotic subregions separated by transport barriers formed by segments of certain stable and unstable manifolds (as well as the regular islands) were identified earlier in~\cite{Bohigas93} and the relevant figure is reproduced in Fig.~\ref{regions}.  Classically, there are eleven subregions, however four of them are related by a reflection symmetry.  Building the symmetry into the quantum mechanics effectively leaves only seven distinct subregions (related regions are denoted by a prime, as in $5$ and $5^\prime$).  The various phase space volumes and connecting fluxes are all known.  As a final note, ahead time is measured in periods of the central closed orbit of the island $\mathcal{A}$ at the mean energy of the wave packets being used.  Roughly then, $t=10$ can be thought of as the time a trajectory crosses back and forth across the width of the system a full ten times.

\section{Classical, semiclassical, and quantum results}
\label{sec-5}

Although the primary goal is to investigate whether HK implementation of semiclassical theory is sufficient to deduce localization information, it is valuable to begin by considering a case in which a classical transport barrier has no crossing flux.  Amusingly, this may lead to the opposite situation of transport suppression.  In this case, quantum propagation may reach classically forbidden subregions by way of tunneling.  It turns out that the zero-flux regime can provide at least a couple of critical pieces of information.  First of all, for quantum transport suppression to be understood properly, it is necessary to know how well a wave packet can be contained within a subregion.  A wave packet does have exponentially suppressed tails that extend beyond whatever region is being considered.  Ideally, their total weight external to their primary subregion should be on a far, far smaller scale than the extent of transport suppression one is trying to observe.  One would thus know in which regimes those tails are safely ignored.  Cross-correlation functions evaluated to intermediate propagation times in a zero-flux regime reflect the measure of the tails, that is at least up to times where tunneling begins to take over.  Secondly, the HK semiclassical propagator can seriously and erroneously magnify highly unstable trajectories leading to inaccurate propagation in strongly chaotic dynamical situations.  This problem tends to grow exponentially with propagation time in chaotic systems.  The zero-flux regime allows one to see rather precisely at which times a particular HK propagation implementation loses accuracy.  There the trajectories associated with the tails abruptly become magnified orders of magnitude beyond their true contributions, and the onset is readily seen.  Increasing the number of trajectories in the implementation should then be seen to delay the onset of errors.

\subsection{The zero-flux regime}

The only zero-flux barriers in the quartic oscillators' dynamics effectively are the boundaries of the regular islands, but they can be used for our purposes.  The idea is to propagate a wave packet initially as well localized as possible within one such island and check its overlap with a wave packet localized as well as possible in its neighboring region.  We choose the initial Gaussian wave packet inside island $\cal A$ and measure its overlap with a final Gaussian wave packet within region $5^\prime$.
\begin{figure}
\subfigure{\includegraphics[width=0.5\textwidth]{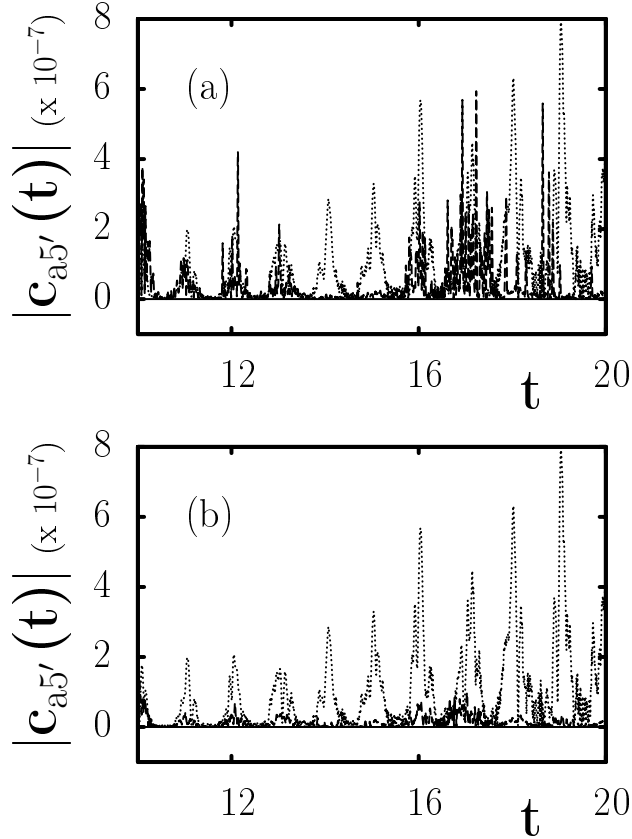}}
\caption{Classical (solid horizontal line), semiclassical (long dashes) and quantum mechanical (dotted line) cross-correlation functions for the regular island with a probe gaussian in the chaotic region for different numbers of trajectories (tr.) in the HK calculations: $2\cdot 10^6$ tr. in (a), $64\cdot 10^6$ tr. in (b).  The classical calculations were performed with $10^4$ trajectories.  The quantum result was obtained with the split-operator fast Fourier transform technique~\cite{Fleck76}.  }
\label{tunnel}
\end{figure}

It is known that for this quartic oscillator system, chaos-assisted tunneling~\cite{Tomsovic94}, which is a specific mechanism of dynamical tunneling~\cite{Davis81}, is present for all the islands.  This leads to an increasing probability for the propagated quantum state to have an overlap in the chaotic sea, but on a very long time scale, much longer than the local mixing time.  Since the HK propagation is not implemented with tunneling trajectories, it should not show this increase at all, should be less than the quantum probability, and should be close to the classical result.  The classical, HK semiclassical, and quantum magnitudes of the cross correlation function are plotted in Fig.~\ref{tunnel}.  It is worth pointing out several features.  Of primary importance is that the scale of the cross-correlation function is $O(10^{-7})$, which is much more than small enough for the purposes of studying suppression of transport ahead.  It is also possible to see the increasing overlap with time due to the tunneling.  Next it is worth commenting on the HK implementation.  Considering the geometry of the semiclassical calculations, it is sufficient to choose a phase space annulus near the boundary of the regular island mainly in the chaotic region for the initial conditions of trajectories in Eq.~(\ref{hk}).  The contribution from the inner part of the wave packet is vanishing as those trajectories remain trapped well within the island.  The upper limit of the ring is taken in such a way that the contribution from all integration points beyond this limit is negligible.  Nevertheless, even with this simplification, $2\times 10^6$ trajectories in the HK calculation is not yet converged to below the tunneling contributions at the intermediate time scale shown.  This behavior is cured by going to $64\times 10^6$ trajectories.  This is a very stringent test in a strongly chaotic system and again well beyond what is necessary for the transport suppression study.  To be a little more specific about the problems that chaotic systems create, when a part of the integration region is chaotic, two features work together to drastically hinder the convergence of the semiclassical integral.  Both originate in the sensitive dependence of the trajectories on their initial conditions. First of all, a set of initial points limited to a small part of the phase space still spread over the entire phase space. This implies a highly oscillatory integrand. Consequently one needs many more initial points than in the integrable case to obtain a sufficient coverage of the final phase space. Secondly, the prefactor in Eq.~(\ref{hk}) is a measure for the stability of the trajectories and it becomes exponentially large in time.  It is the contributions of a locally dense set of trajectories with various phases that lead to cancellation of most of this magnitude.  Again, it requires a number of sampling points being of the same order of magnitude to preserve the norm.  Therefore, to improve the convergence of the integral over the annulus we employed the time-integration method presented in~\cite{Elran99}. Also a cutoff value for the prefactor was used, as proposed in~\cite{Elran99, Kay94c}. In this way about $10\%$ of the trajectories were discarded after a certain time.

\subsection{Partial transport barriers}

Moving on to the transport with non-vanishing fluxes, the time-averaged transition probability is the physical quantity of interest.  Several different examples are calculated that illustrate crossing, one, two or more barriers.  In each case, the initial wave packet is the same, in the region $5^\prime$ as indicated in Fig.~\ref{regions}, and the mean energy is in the neighborhood of the
1000th excited energy eigenstate. 
\begin{figure}
\includegraphics[width=0.5\textwidth]{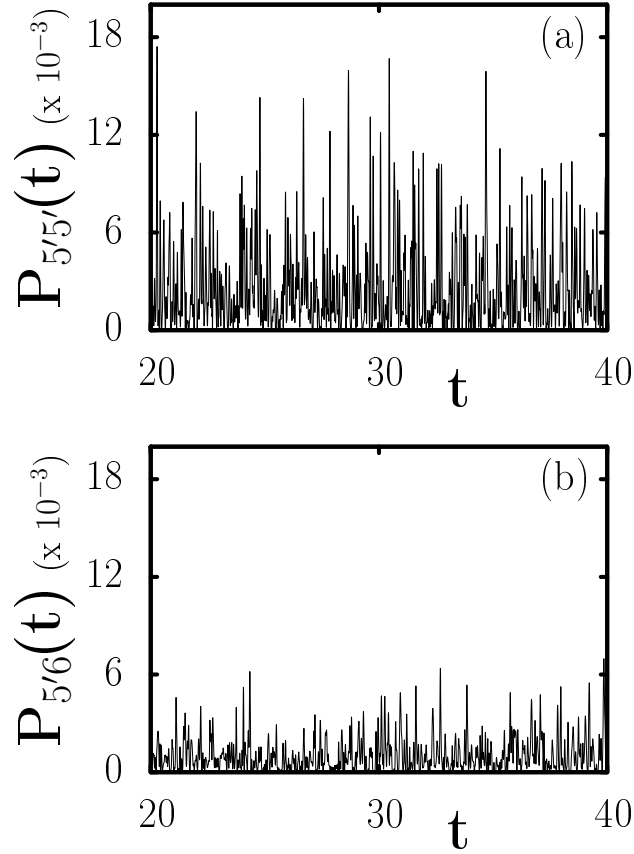}
\caption{Quantum mechanical transition probabilities in the zero barrier case 
(a) and for one barrier (b) at long propagation times.  For the zero barrier case, both wave packets are in region $5^\prime$ and for the one barrier case, the final wave packet is placed in region $6$ (the regions are shown in Fig.~\ref{regions}).  The initial wave packet width is $\gamma_{5^\prime} = 3.3$. }
\label{qm}
\end{figure}
It turns out to be extremely convenient for comparison purposes to arrange that the smoothed local spectra of all the final wave packets considered have the same shape and width.  In this way, the integral over the local spectra in Eq.~(\ref{C_pre}) is identical in all cases.  The criterion needed to ensure this is that the short time decay of the corresponding autocorrelation functions, in particular its absolute value, has to be equal.  That simplifies the ratio of any two time-averaged transition probabilities to be
\begin{equation}
\label{ratio}
\frac{{\cal P}_{jk}}{{\cal P}_{jl}} = \frac{ \Delta_{jk}({\bf \Lambda})}{ \Delta_{jl}({\bf \Lambda})}\frac{f_l}{f_k}
\end{equation}
In order to arrange an appropriate initial decay behavior, it is sufficient to use the frozen Gaussian approximation~\cite{Heller81}, assuming that on this time scale the width of the Gaussian changes only slightly.  Expanding the potential up to first order, solving the canonical equations, and using the analytic result for Gaussian integrals with equal widths yields
\begin{equation}
P_{kk}(t) \simeq \exp \left[-\frac{\gamma_k}{2}(\mb{V}_{\mb{q}_0} t^2 -\mb{p}_0 t)^2- \frac{1}{2\hbar^2\gamma_k}\mb{V}_{\mb{q}_0}^2 t^2\right],
\label{abs-acf}
\end{equation}
where $\mb{V}_{\mb{q}_0}$ denotes the vector of the partial derivatives of the potential with respect to the positions at the initial point. As only the short time behavior is relevant, Eq.~(\ref{abs-acf}) can be expanded to the quadratic time terms:
\begin{equation}
P_{kk}(t) \approx 1-\left(\frac{\mb{p}_0^2\gamma_k}{2}+\frac{1}{2\hbar^2\gamma_k}\mb{V}_{\mb{q},0}^2 \right) t^2 \equiv 1-\chi_i t^2.
\end{equation}
Assuming the mean energies of two wave packets are equal, then the smoothed local spectra cover the same energy range if $\chi_j=\chi_k$.  Given the width of one wave packet, say the initial one $\gamma_j^{-1}$, the width of the other, $\gamma_k^{-1}$, is found by the relations
\begin{equation}
\gamma_k=\frac{\gamma_j \mb{V}_{\mb{q}_{0,k}}^2}{ \mb{V}_{\mb{q}_{0,j}}^2 + \hbar^2 \mb{p}_{0,j}^2 \gamma_j^2 } 
\end{equation}
if $\mb{p}_{0,k}^2=0$ and
\begin{equation}
\gamma_k=\frac{1}{2\gamma_j\hbar^2\mb{p}_{0,k}^2 } \left(\mb{V}_{\mb{q}_{0,j}}^2 + \hbar^2\gamma_j^2 \mb{p}_{0,j}^2 + \sqrt{\left(\mb{V}_{\mb{q}_{0,j}}^2 + \hbar^2\gamma_j^2 \mb{p}_{0,j}^2\right)^2- 4 \gamma_j^2 \hbar^2\mb{V}_{\mb{q}_{0,k}}^2\cdot \mb{p}_{0,k}^2} \right) 
\end{equation}
otherwise. 

The transition probability for a case with no barrier crossing can be arranged by placing the final wave packet into the same region as the initial wave packet such that their initial overlap is negligible.  To arrange a one barrier crossing example, the final wave packet is placed into the appropriate neighboring region.  This leads to results of the kind shown in Fig.~\ref{qm}.  There is a visible decrease in the locally averaged transition probability for the one-barrier case relative to the zero-barrier case. 

For a clearer visualization of the information about transport suppression, the transition probability is smoothed with a Gaussian window function [as opposed to the ``square'' window used in Eq.~(\ref{Delta_orig})] given by
\begin{equation}
{\cal P}_{jk} (t_0;\zeta) = \frac{1}{\sqrt{2\pi}\zeta}\int\limits_{-\infty}^{+\infty} \di t \, \eh ^{-(t-t_0)^2/(2\zeta^2)} P_{jk}(t),
\end{equation}
where $\zeta$ is a smoothing parameter. The value of $\zeta$ is ideally chosen to be as small (narrow) as possible while still being capable of smearing out the rapid transition probability oscillations.  Although the smoothing is not infinitely wide, it nevertheless should approach the ratio given in Eq.~(\ref{ratio}).

\begin{figure}
\includegraphics[width=0.5\textwidth]{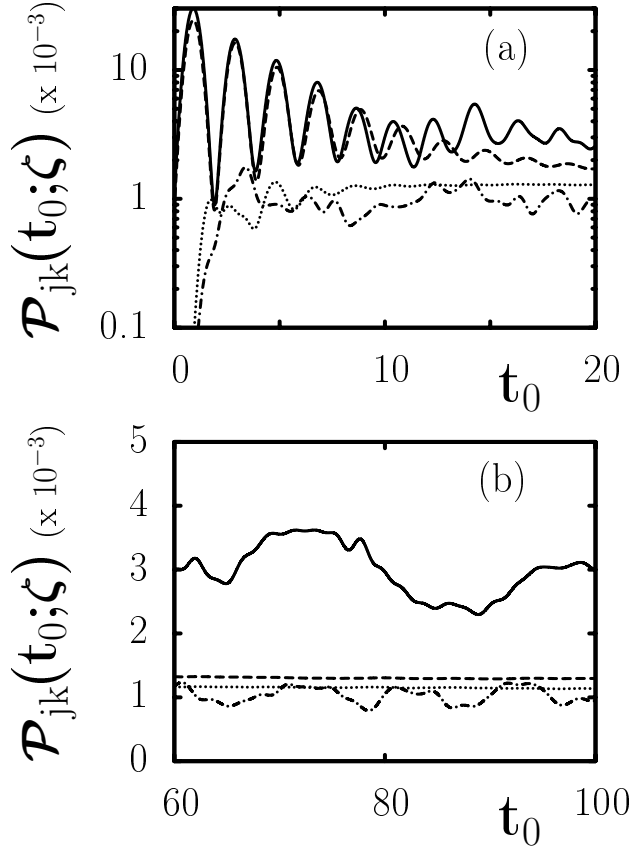}
\caption{Locally time-smoothed quantum and classical transition probabilities for zero ($j=5',k=5'$) and one barrier ($j=5',k=6$). The quantum zero-barrier case is represented by the solid line, the classical zero-barrier case by the dashed line, the quantum one barrier case by dash-dots, and finally the classical one-barrier case by the dotted line.  The smoothing parameter $\zeta=0.35$ in (a) and $0.8$ in (b) and the results shown are obtained with $10^6$ trajectories.  Note that the two classical cases are very close to each other at longer times whereas the two quantum cases are not. Also note the logarithmic scale in (a).}
\label{qm-cl}
\end{figure}
\begin{figure}
\includegraphics[width=0.5\textwidth]{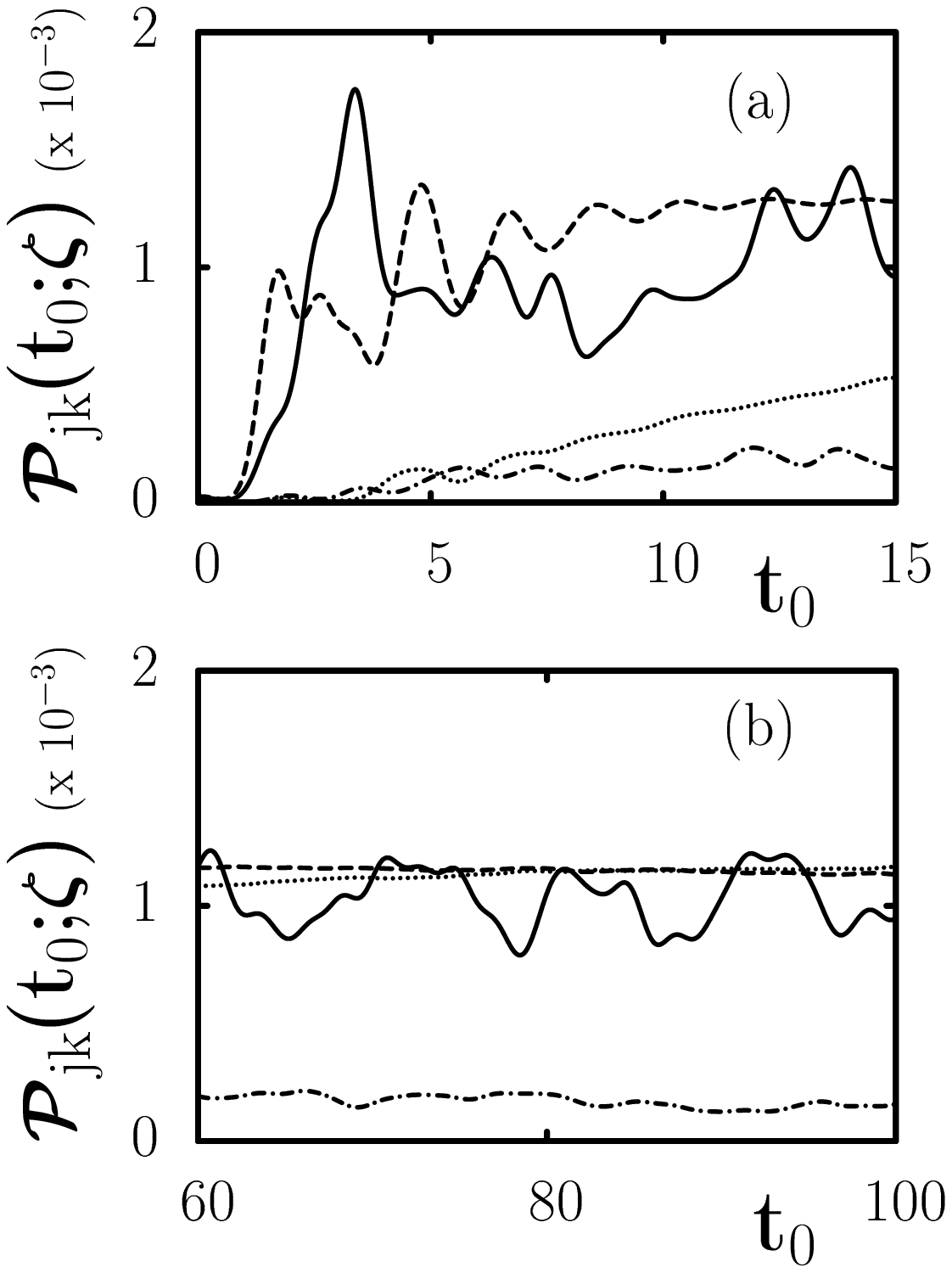}
\caption{Locally time-smoothed quantum and classical transition probabilities for one ($j=5',k=6$) and three barriers ($j=5',k=2$). The quantum one-barrier case is represented by the solid line, the classical one-barrier case by the dashed line, the quantum three-barrier case by dash-dots, and finally the classical three-barrier case by the dotted line.  The smoothing parameters and numbers of trajectories are identical to Fig.~\ref{qm-cl}.  Here the two classical cases take much longer to approach each other, but eventually do.  On the other hand, the ratio of the values of the two quantum cases are far from unity indicating the extra localization due to three barriers rather than one.}
\label{qm-cl2}
\end{figure}
In Fig.~\ref{qm-cl}, the quantum and classical time-smoothed transition probabilities are compared for the zero-barrier and one barrier cases.  The shorter time dynamics in the upper panel illustrates the transient behavior and the bottom panel illustrates the long time average properties.  In the transient behavior, the effect of the barrier is to delay the onset of the transition probability for both quantum and classical behaviors.  In the theory given in~\cite{Bohigas93}, the one-barrier quantum and classical behaviors are both predicted to have a linear increase initially, albeit with superposed oscillations.  It is this time regime in which the HK propagation needs to be accurate if it is to be able to contain the information about long-time localization.  At long times, the classical results cannot show any localization distinction and the ratio of the value of those two curves must approach unity by the time that mixing becomes complete.  One sees clearly that this is not what happens at long times for the quantum curves and that the localizing effects due to the barrier prevents their ratio from approaching unity.  Crossing more barriers only increases these effects.  In Fig.~\ref{qm-cl2}, a wave packet is placed in region 2 [see Fig.~\ref{regions}], which means that three barriers are effectively being crossed, and the transition probabilities are compared with the zero-barrier case shown in the previous figure.  At short times, the onset delay is much clearer, but nevertheless, remarkably similar in the quantum and classical cases.  Here, according to the approach of~\cite{Smilansky92}, the initial increase is expected to be cubic in time.  This follows by the tri-diagonal nature of the flux matrix, $\bf A$, for this case and the expansion
\begin{equation}
{\bf P}(t)  = \exp(-\mb{A}t){\bf P}(0) = \left(\mb{1}-\mb{A}t+\frac{1}{2}\mb{A}^2t^2 -\frac{1}{6}\mb{A}^3t^3\right){\bf P}(0)+\mathcal{O}(t^4)
\label{master}
\end{equation}
Until the $\bf A$ matrix is cubed, the element of $\exp(-{\bf A} t)$ connecting one region to another across three barriers is null.  In the long time dynamics, again the two classical transition probabilities approach each other, but it does take more time, and that is just as expected.  The ratio of the local mean values of the quantum curves is again far from unity, which implies further localization also as expected. 

\begin{figure}
\includegraphics[width=0.5\textwidth]{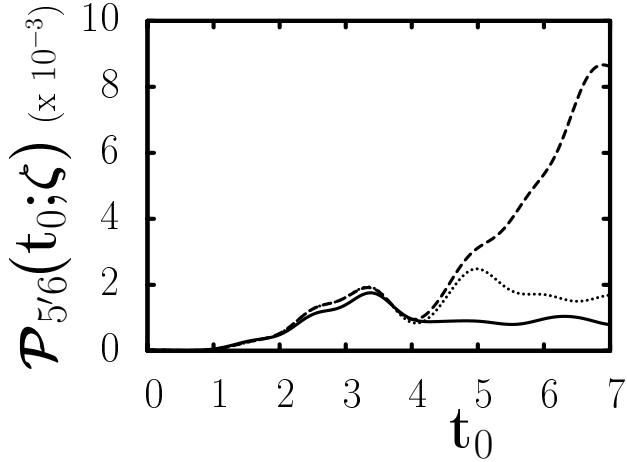}
\caption{Comparison of the semiclassically and quantum mechanically calculated transition probability. Shown is the one-barrier case in the previous two figures: quantum result (solid line) versus semiclassical result with $10^6$ trajectories (dashed line) and semiclassical result with $10^7$ trajectories (dotted line).  The convergence for longer times with greater numbers of trajectories is clear.}
\label{hk-qm}
\end{figure}
Next, consider the time-dependent semiclassical description of the zero and one barrier dynamics.  First of all, it is important to understand the convergence behavior of the HK propagator given the presence of strongly chaotic dynamics.  As an example, the semiclassical transition probability for the one-barrier case is compared with the quantum transition probability for different numbers of sampling points in Fig.~\ref{hk-qm}.  The numerical integration was performed with the importance sampling Monte-Carlo method~\cite{Kluk86}.  Note that the time integration method was not applied here, but a cutoff for the prefactor was used, too.  It leads to discarding around $15\%$ of the trajectories at the final integration time.  Up to time $t=4$, there is no significant difference between the semiclassical curves for a different number of sampling points. After that time, it is necessary to use the larger number of trajectories, $10^7$, to achieve reasonably close convergence to the quantum result, although a complete convergence is not yet attained.  

\begin{figure}
\includegraphics[width=0.5\textwidth]{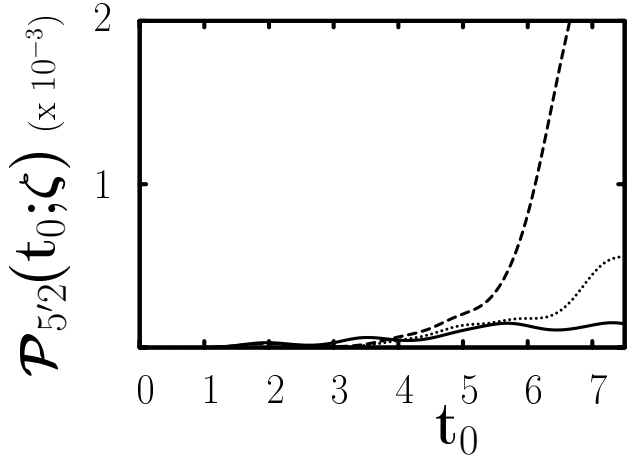}
\caption{Comparison of the semiclassical and quantum mechanically calculated transition probability for the three barrier case: quantum result (solid line) versus the semiclassical result with $10^6$ trajectories (dashed line) and the semiclassical result with $10^7$ trajectories (dotted line).}
\label{hk-fig}
\end{figure}
The convergence with the greater number of trajectories turns out to be sufficient for the demonstration of localization due to transport barriers, but having sufficient trajectories is absolutely critical.  
In Fig.~\ref{qm-cl2}, it is seen that the transient rise to the long time average occurs mostly in the time range $1\le t\le 4$ for the one-barrier case and $3\le t\le 7$ for the three-barrier case. It is this time range from which the long time localization can be deduced. The greater number of trajectories does give convergence there as seen also in Fig.~\ref{hk-fig} for the three barrier case. 
However, a simple renormalization of the wave function for low numbers of trajectories as done in~\cite{Yoshida04} would not incorporate this behavior and would be insufficient for understanding localization.

We now arrive at the crux of the matter, i.e.~whether a realizable implementation of the HK propagator can contain the information necessary to deduce localization properties and whether that information can be extracted.  To understand the results for the one-barrier case quantitatively, consider first the ratios of the long-time averages of the transition probabilities in the quantum and classical cases. The exact classical ratio always approaches unity; in our numerical results (see Fig.\ \ref{qm-cl}) a close value of ${\cal P}_{56}/{\cal P}_{55}=0.89$ is found.  In the next paragraph, the many trade-offs made in order to set up the test given here is addressed and how that can lead to the classical result not being exactly unity.  Given this classical result, roughly speaking, a 10\% variation due to all the competing conditions is about the best accuracy that can be expected.  Back to the quantum result, which reflects the localization of interest: the further from unity, the greater the localization.  For the one-barrier to zero-barrier ratio, the quantum calculations give for the ratio $0.34$, which is clearly well away from unity to the level of variation that can be controlled, and it reflects the effects of the transport barrier.  The next step is to find out how this compares to the semiclassical predictions. There are two ways to do this, and they only rely on the intermediate time dynamics to work.  The first possibility is to use the perturbative analytical results incorporating the measured classical fluxes, relative volumes of the subregions, and the mean excitation energy (in order to obtain the smoothed level density), which can all be obtained from \cite{Bohigas93}, in Eq.\ (\ref{ratio}) together with Eqs.\ (\ref{lambda},\ref{Delta}).  Note that if the system is not in the perturbative regime, the Master equation approach would have to be used.  The other way possible in the perturbative regime includes the usage of the results from the HK propagation. We have shown that the HK method is capable of reproducing the behavior in the transient regime correctly, although great care must be exercised in order to ensure that enough trajectories are included in the implementation.  To obtain information about the long time behavior, one can deduce numerically an approximate slope from this regime (see again Fig. 7), which corresponds to the linear coefficient of the transition probability in Eq.\ (\ref{ampli}). This again can be used to extract $\Lambda_{56}$ respectively $\Delta_{56}$. Together with $\Delta_{54}$ from the perturbative analytical results one obtains $\Delta_{55}$ and furthermore the ratio.  These two methods give $0.38$ and $0.30$, respectively, and bracket the actual quantum result. They are within roughly 10\% of the quantum result.  To the level that could have been expected under the conditions of the calculation, it indeed was possible to extract the information from the results using the HK propagator.

Finally, it is worth highlighting explicitly and all together in one location, the nature of a number of the desires and approximations required to obtain the results in the above paragraph.  It is desirable to remain on an energy surface that is excited sufficiently that semiclassical approximations are valid.  In particular, in addition to the usual semiclassical considerations, localizing a wave packet within a particular region in this problem improves as the ratio $E^{3/4}/\hbar$ increases.  The smaller a region or the more filamentary, curved, or possessing of small regular islands, the more difficult it is to locate a wave packet wholly within that region and have the tails avoid regular islands.  This argues for increasing the energy or conversely shrinking $\hbar$.  Working in the opposite direction is the desire that the example not be so excited as to be way beyond the regime that might be of interest in physical chemistry.  That desire also has the added benefit of keeping the transition parameters within the perturbative regime for the particular quartic oscillators studied ($\lambda=0.35$).  Much higher in energy than the example given here and that would no longer have been possible.  Making the entire enterprise more complicated is the use of wave packet widths such that their strength functions had the same energy spread in order to simplify the interpretation of the final results.  The end result is that the wave packet tails did cross subregional boundaries, including regular islands.  This explains the classical ratio above not approaching unity, but rather 0.89.  Similarly, competing effects are going on in the quantum case where the agreement is quite good, but at least part of the quality of the agreement is likely due to canceling opposing effects of wave packet tails crossing boundaries between two chaotic subregions (inhibiting a finding of localization) and the effects of tails crossing into a regular region (enhancing a finding of localizing effects).

\section{Conclusions and Outlook}

The principal concern in this paper is whether an initial value representation of the semiclassical propagator heavily relied upon in molecular and atomic contexts, the Herman-Kluk propagator, can be implemented well enough to capture the information necessary to understand subtle long-time quantum transport effects, such as wave packet localization.  The ``freezing out'' of wave packet exploration or transport of energy is potentially extremely important.  In this quest, for many reasons the possibility of success can only arise if prior analysis allows one to deduce the long-time effects from much shorter time dynamics.  Not only does the Herman-Kluk propagator fundamentally lose accuracy with increasing time scales, an implementation goes more and more beyond practical computational limits, and in addition, is made more difficult as a system's dynamics becomes more strongly chaotic and as the number of degrees-of-freedom increases.   As was shown previously~\cite{Bohigas93}, the wave packet localization induced by classical transport barriers such as cantori or crossings of stable and unstable manifolds provides one arena in which the relatively short-time and long-time behaviors are intimately linked and understandable.  

A system of two coupled quartic oscillators was investigated as a simple, convenient, previously-explored example of coupled anharmonic oscillators that could be expected to contain similar dynamical features found in models of molecular and atomics systems (although perhaps they are more strongly chaotic than typically found in these latter models and thus more challenging in that sense), and in particular, they display a variety of classical transport barriers.  Another helpful feature is the existence of islands of regular motion.  They could be used effectively as zero-flux transport barriers, which in turn provided a rather exacting test of the HK propagator implementation.  Although, dynamical (actually, chaos-assisted) tunneling is present, it is weak enough to be orders of magnitude smaller than the effects of the classical transport barriers of interest.  By leaving out tunneling trajectories in the HK implementation, convergence could be imposed to below the importance of the tunneling contributions seen in the full quantum dynamics.

At least in the quartic oscillators, a fairly strongly chaotic two-degree-of-freedom system, it turned out to be possible to obtain accurate propagation throughout the short to intermediate times needed to deduce the long-time wave packet localization.   The amount of localization deduced from the HK calculation matched the actual quantum result to within 10\%.   Perhaps, greater implementation efficiency and less unstable dynamics would allow for the treatment of systems with a few more degrees-of-freedom.  It would be worthwhile to understand where the upper bound in treating more degrees-of-freedom is.  

As a final point, there has been a Wigner function based study of 
recurrences using the Pullen-Edmonds Hamiltonian, and it has revealed the usefullness 
of classical trajectory approaches in another mixed phase space system 
\cite{Zdanska01}.  Therefore, there is every reason to expect that it should be possible to study localization semiclassically in non-energy-scaling, a-bit-more-realistic systems.

{\bf Acknowledgments:} The authors acknowledge fruitful discussions with and support from Denis Ullmo.  They also benefited from computational resources by the Zentrum f\"ur Informationsdienste und
Hochleistungsrechnen (ZIH) of the Technische Universit\"at Dresden.  FG and CMG gratefully acknowledge financial support by the Deutsche Forschungsgemeinschaft (GR 1210/4-1).  ST gratefully acknowledges support from the Max-Planck-Institut f\"ur Physik komplexer Systeme and the US National Science Foundation grant  No.~PHY-0555301.


\end{document}